\documentclass[12pt,aps,print,showpacs]{revtex4-1}
\usepackage{amsfonts}
\usepackage{amsmath}
\usepackage{amssymb}
\usepackage{bbold}
\usepackage{braket}
\usepackage{graphicx}
\usepackage{epstopdf}
\usepackage{subfigure}
\usepackage{hyperref}

\begin{document}

\title{Comparing classical and quantum PageRanks}
\author{T. Loke$^{1}$}
\email{21065783@student.uwa.edu.au}
\author{J. W. Tang$^{1,2}$}
\author{J. Rodriguez$^{1}$}
\author{M. Small$^{3}$}
\author{J. B. Wang$^{1}$}
\email{jingbo.wang@uwa.edu.au}
\affiliation{{$^{1}$School of Physics, The University of Western Australia, WA
6009, Australia}\\
{$^{2}$School of Physics, Nanjing University, Jiangsu, China}\\
{$^{3}$School of Mathematics and Statistics, The University of Western Australia, WA
6009, Australia}}

\begin{abstract}
Following recent developments in quantum PageRanking, we present a comparative analysis of discrete-time and continuous-time quantum-walk-based PageRank algorithms. For the discrete-time case, we introduce an alternative PageRank measure based on the maximum probabilities achieved by the walker on the nodes. We demonstrate that the required time of evolution does not scale significantly with increasing network size. We affirm that all three quantum PageRank measures considered here distinguish clearly between outerplanar hierarchical, scale-free, and Erd\"os-R\'enyi network types. Relative to classical PageRank and to different extents, the quantum measures better highlight secondary hubs and resolve ranking degeneracy among peripheral nodes for the networks we studied in this paper. 
\end{abstract}

\maketitle	

\section{Introduction}
\label{sec:Introduction}

Characterising the relative importance of nodes in a graph is a key element in network analysis. A ubiquitous application of such centrality measures is Google's PageRank algorithm \cite{brin2012reprint, page1999pagerank}, whereby the World-Wide Web (WWW) is considered as a network of webpages (nodes) connected by hyperlinks (directed edges) between them. By ranking each webpage according to its PageRank centrality, the search engine's results are ordered based on their approximated quality.

There has been recent interest in formulating a quantum version of PageRank. Since the intuition behind Google's PageRank is a classical ``random surfer'' crawling the WWW, a quantum walker traversing the associated directed network can be expected to provide an analogous measure of PageRank. As the quantum analogue of classical random walks, quantum walks serve as building blocks for quantum algorithms that can outperform their classical counterparts \cite{ambainis2003quantum}. It is thus interesting to study whether their quantum mechanical properties afford an advantage over Google's classical PageRank algorithm.

Paparo \textit{et al.} \cite{paparo2012google} and S\'anchez-Burillo \textit{et al.} \cite{sanchez2012quantum} have separately proposed two quantum PageRank measures. The former is based on a discrete-time quantum walk (DTQW), whereas the latter uses a continuous-time quantum walk (CTQW). While quantum walks on arbitrary undirected graphs have been well defined, extending this framework to include directed quantum walks is non-trivial due to the requirements of unitarity and reversibility of the walk \cite{montanaro2005quantum}. To deal with this difficulty, the discrete-time quantum PageRank uses a non-fully-directed but unitary walk; whereas the continuous-time algorithm forgoes unitarity in using an open-system quantum walk.

In \cite{paparo2013quantum}, Paparo \textit{et al.} performed further analysis of their proposed quantum PageRank on complex networks, specifically on hierarchical graphs, directed scale-free graphs, and Erd\"os-R\'enyi random graphs. The quantum PageRank algorithm not only distinguished clearly between the three graph classes, but also exhibited distinct characteristics in terms of highlighting secondary hubs and lifting the degeneracy of low-lying nodes. While it displayed a smoother power law behaviour on scale-free networks, it was more sensitive to coordinated attacks on hubs than the classical PageRank algorithm.

Nevertheless, the number of time steps required for the underlying discrete-time quantum walk to yield a reliable quantum PageRank is yet to be considered. We seek to address this by investigating the oscillatory nature of the walker's probability amplitudes across nodes in the network. Such a consideration is worthwhile should an efficient quantum-system-based implementation of the PageRank scheme become realisable.

The open-system-quantum-walk-based PageRank in \cite{sanchez2012quantum} modelled directionality as the walker's non-unitary interaction with the environment. Similar to the discrete-time case, the open-system PageRank lifted classical rank degeneracy of lowly connected nodes, whilst preserving identification of the most central nodes. By extension, it is useful to ascertain whether the other characteristics found in \cite{paparo2013quantum} for the discrete-time quantum PageRank are reflected in the open-system scheme.

In this article, we largely follow the analysis in \cite{paparo2013quantum}, but extend it in three ways. Firstly, we consider the time-scale involved for discrete-time-quantum-walk-based PageRank. For the network types considered here, we gauge a suitable number of time steps for the walker's evolution after which reliable PageRanks can be obtained. We propose such an upper bound that does not scale significantly with increasing network size.

Secondly, rather than taking the time average of the walker's probability distribution, we propose an alternative indicator of PageRank based on the maximum probability amplitude achieved by the walker on each node. This has previously been studied as a centrality measure on undirected graphs in \cite{berry2010quantum}.

Thirdly, we concurrently analyse an open-system-based PageRank algorithm. In our comparative study of three quantum PageRank schemes, we discuss their relative performance in extracting practically useful information about the networks under consideration. This provides a better understanding of each scheme as tools for quantum-walk-based complex network analysis. Our results suggest that as per classical PageRank, quantum PageRanking distinguishes clearly between the outerplanar hierarchical, scale-free, and Erd\"os-R\'enyi network families. While the quantum measures pick out more secondary hubs and remove degeneracies among low-lying nodes \cite{paparo2013quantum, sanchez2012quantum}, each exhibits such quantum advantage to different extents.

This article is organised as follows: Section \ref{sec:Theory} outlines the theoretical framework underlying the classical and quantum PageRank algorithms. In Section \ref{sec:Results}, we present our numerical results for the algorithms on three types of directed networks, namely outerplanar hierarchical, scale-free, and Erd\"os-R\'enyi networks. We continue our comparative analysis on the algorithms in terms of secondary hub resolution on scale-free networks, localisation-delocalisation of the walker, and power law behaviour on scale-free networks. Finally, Section \ref{sec:Discussion and conclusions} contains discussion and conclusions.

\section{Theory}
\label{sec:Theory}

\subsection{Classical PageRank}
Google's PageRank algorithm is a variant of eigenvector centrality. The PageRank vector $I_{cl}$ is given by
\begin{equation}
\label{eq:CPR}
GI_{cl} = I_{cl}
,
\end{equation}
where $G$ is the Google matrix, defined as 
\begin{equation}
\label{eq:matG}
G:= \alpha E + \frac{(1-\alpha)}{N}\textbf{1}
.
\end{equation}
Here $N$ is the number of nodes in the network, $E$ is a (patched) connectivity matrix, $\alpha$ is the damping parameter (typically $\alpha=0.85$), and $\textbf{1}$ is the matrix of all ones. Intuitively, the second term represents the possibility of the walker randomly hopping to any other node in the network \cite{page1999pagerank}.

Define the connectivity (or adjacency) matrix $C$ of the network as $C_{jk}=1$ if there is an edge from $k$ to $j$, and $C_{jk}=0$ otherwise. To obtain the patched $E$, $C$ is modified such that each column $k$ containing all zeroes (corresponding to a node $k$ with zero out-degree) is replaced by a column with all entries set to $\frac{1}{N}$.  The remaining columns corresponding to nodes with outgoing link(s) are normalised to sum to one by dividing by the out-degree of the node.
Denote the out-degree of a node $k$ by $D_k$, with $D_k=\sum_j C_{jk}$. Mathematically, $E$ is then
\begin{equation}
E_{jk} =
\begin{cases} \frac{1}{N} &\mbox{if } D_k = 0 \\ 
\frac{C_{jk}}{D_k} & \mbox{if } D_k \neq 0
\end{cases}
\label{eq:matE}
\end{equation}
and is in general column stochastic.

\subsection{Szegedy-Google PageRank via discrete-time quantum walk}
Szegedy's formalism of the discrete-time quantum walk is a quantisation of the Markov chain corresponding to a classical random walk \cite{szegedy2004quantum, szegedy2004spectra, aharonov2001quantum, childslecturenotes}. Classically, for an $N$-node graph, such a process is described by an $N$-by-$N$ matrix $P$ of transition probabilities, where each entry $P_{jk}$ denotes the transition probability from node $k$ to node $j$.
Szegedy's walk takes place on the Hilbert space $\mathcal{H}^{N^{2}} = \mathcal{H}^{N} \otimes \mathcal{H}^{N}$. This space is the span of all vectors $\ket{j,k}$, where each vector represents a directed edge in the graph from node $j$ to node $k$.

First we define the state vector
\begin{equation}
\label{eq:SG1}
\begin{split}
\ket{\psi_{j}}&:=\ket{j}\otimes\sum_{k=1}^{N}\sqrt{P_{kj}}\ket{k}\\
&=\sum_{k=1}^{N}\sqrt{P_{kj}}\ket{j,k}
\end{split}
\end{equation}
for each node $j=1,\ldots,N$ of the graph. This represents a superposition of edge states $\ket{j}_1\ket{k}_2$ outgoing from the $j$th vertex, weighted by $P$.  The reflection operator is given by
\begin{equation}
\label{eq:SG2}
\hat{\Pi}:=\sum_{j=1}^{N}\ket{\psi_{j}}\bra{\psi_{j}}
,
\end{equation}
and
\begin{equation}
\label{eq:SG3}
\hat{S}:=\sum_{j,k=1}^{N}\ket{j,k}\bra{k,j}
\end{equation}
is the swap operator. Then a step of the quantum walk is the unitary operator
\begin{equation}
\label{eq:SG4}
\hat{U}:= \hat{S}(2\hat{\Pi}-\hat{\mathbb{1}}),
\end{equation}
whereas a two-step evolution operator takes the form
\begin{equation}
\label{eq:SG5}
\hat{U}^2:= (2\hat{S}\hat{\Pi}\hat{S}-\hat{\mathbb{1}})(2\hat{\Pi}-\hat{\mathbb{1}}).
\end{equation}

As proposed in \cite{paparo2012google}, using the Google matrix $G$ as the stochastic matrix $P$ implements a quantum version of the classical PageRank algorithm. Unitarity of the quantum walk is maintained since $G$ is stochastic, moreover information on the directionality of the network is preserved in $G$.

The corresponding quantum walk is initialised as
\begin{equation}
\label{eq:SG6}
\ket{\psi_0}=\frac{1}{\sqrt{N}}\sum_{j=1}^{N}\ket{\psi_j}
\end{equation}
that is, an equal superposition across all nodes, but weighted among the edge states at each node by $G$.
Taking $\hat{U}^2$ as the discrete time evolution operator of the walk, the instantaneous quantum PageRank is then
\begin{equation}
\label{eq:SG7}
I_q(P_i,t)=\bra{\psi_0}\hat{U}^{\dagger 2t}\ket{i}_2\bra{i}\hat{U}^{2t}\ket{\psi_0}
,
\end{equation}
which is just the walker's probability distribution of over the $P_i$ pages in the network after $t$ time steps. This value does not converge in time to any stationary distribution due to the unitarity and reversibility of the quantum walk operator defined by Eq. \eqref{eq:SG4}.

Since a quantum PageRank measure must provide a unique ranking to each node in the graph, Paparo \textit{et al.} define it as the walker's time-averaged probability distribution:
\begin{equation}
\label{eq:SG8}
I_{TA}(P_i):=\left\langle I_q(P_i,t) \right\rangle=\frac{1}{t_{max}}\sum_{t=0}^{t_{max}-1}I_q(P_i,t),
\end{equation}
which converges for large enough $t_{max}$. This will be referred to as the time-averaged (TA) PageRank measure in this paper.

We propose an alternative PageRank measure based on the peak probability of finding the walker on the node. We use the maximum $I_q(P_i,t)$ reached after $t_{max}$ to be the quantum PageRank of a node: 
\begin{equation}
\label{eq:SG9}
I_{P_{max}}(P_i):=\max \lbrace I_q(P_i,t) : 1 \leq t \leq t_{max}, t \in \mathbb{Z}  \rbrace
.
\end{equation}

We seek to gauge a suitable time-scale $t_{max}$ based on the oscillatory evolution of $I_q(P_i,t)$ according to Eq. \eqref{eq:SG7}. First we apply $t=500$ time steps of $\hat{U}^2$ onto the initial state \eqref{eq:SG6}. Performing a Fourier transform on the time series $I_q(P_i,t)$ yields a power spectrum of the oscillation frequencies present in it. We define $\omega(P_i)$ to be the lowest frequency present above noise using a threshold of 10\% of the highest peak in the power spectrum \cite{berry2010quantum}. The ``period'' of $I_q(P_i,t)$ is then $T_q(P_i)=\frac{2\pi}{ \omega(P_i)}$. In general, each node $i$ in the network, corresponding to page $P_i$, has a different period $T_q(P_i)$.

Denote the mean period of all nodes as $\langle T_{q}^{all} \rangle$:
\begin{equation}
\label{eq:Tq}
\langle T_{q}^{all} \rangle:= \frac{1}{N} \sum_{i=1}^N T_q(P_i)
.
\end{equation}

Let $\langle T_{q}^{5} \rangle$ be the mean period of the five nodes whose instantaneous quantum PageRanks $I_q(P_j,t)$ reach the highest peak values within their respective periods $ 1 \leq t \leq T_q(P_j), t \in \mathbb{Z}$.

Following the above steps, we compute $\langle T_{q}^{all } \rangle$ and $\langle T_{q}^{5}\rangle$ for the directed network families relevant to this study, namely outerplanar hierarchical, scale-free, and Erd\"os-R\'enyi networks with sizes $N=32, 54, 128, 256, 512$ nodes. We use an ensemble of ten scale-free and Erd\"os-R\'enyi random networks for each $N$, generated using NetworkX \cite{schult2008exploring}. For each Erd\"os-R\'enyi network here and throughout this article, the probability for edge creation is set to $p=0.07$. We use $t_{max}=2 \langle T_{q}^{5} \rangle$ as the required time-scale for our PageRank analyses, reasoning that the periods of the most central nodes should figure more strongly over those of the peripheral nodes in determining the general time-scale for each network.

\begin{figure}[tbh]
	\centering
	\vspace{0cm} \hspace{0cm} \scalebox{0.46}{\includegraphics{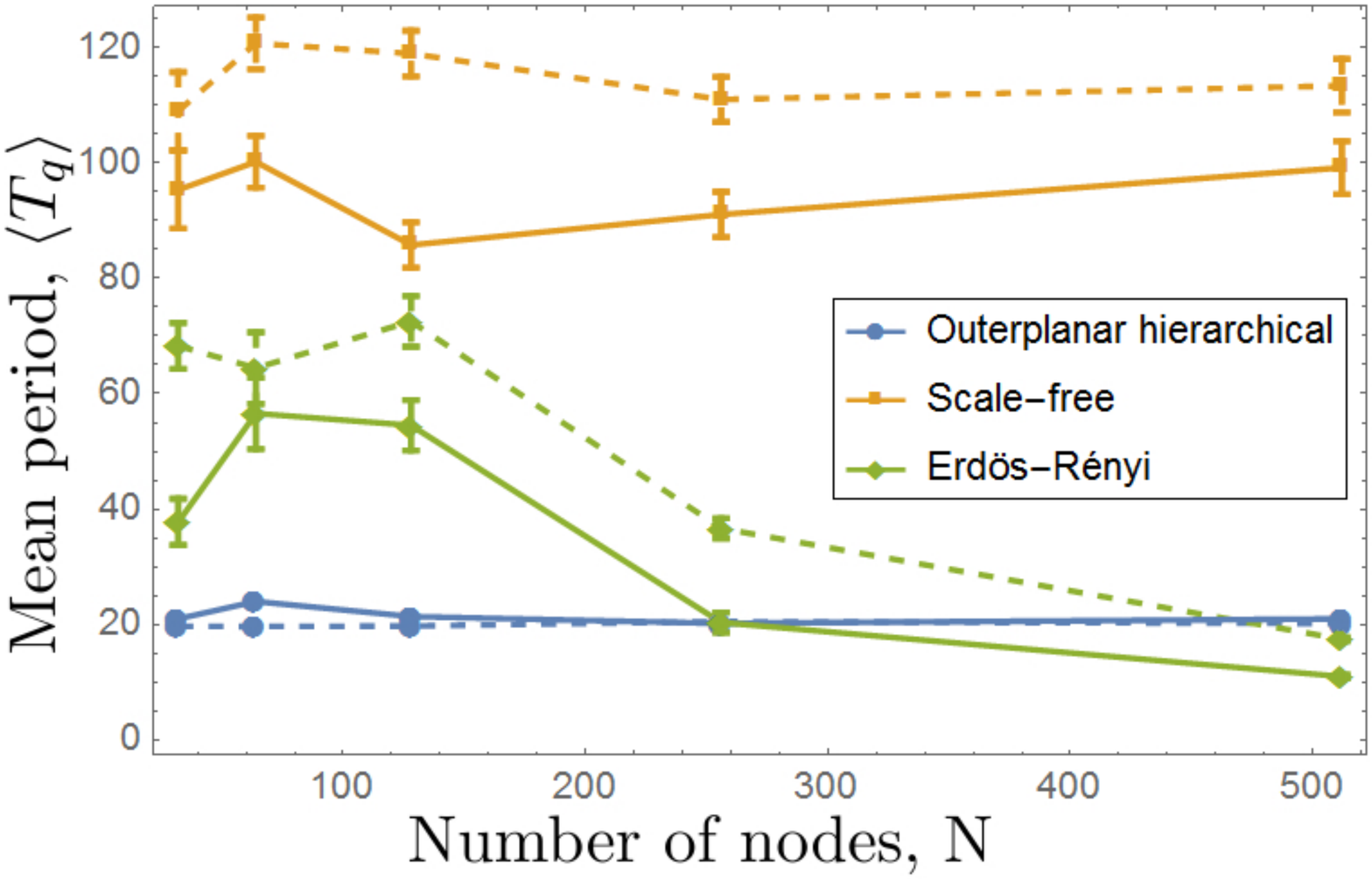}}
\caption{\textbf{Scaling of mean periods with network size for \boldmath$N=32, 64, 128, 256, 512$ nodes.} Mean periods $\langle T_{q}^{5} \rangle$ (solid) and $\langle T_{q}^{all} \rangle$ (dashed) for outerplanar hierarchical, scale-free, and Erd\"os-R\'enyi networks are plotted in blue, orange, and green respectively. For both the scale-free and Erd\"os-R\'enyi random networks, an ensemble of ten graphs is used for each $N$. Each error bar corresponds to the standard error of the mean of the ten $\langle T_{q} \rangle$ values from each ensemble.}
\label{fig:I-T}
\end{figure}

Numerical results are shown in Table \ref{table:T}, and Figure \ref{fig:I-T} plots the scaling of the mean periods with network size. Our results suggest that $t_{max} = 2 \langle T_{q}^{5} \rangle$ does not scale linearly upward with $N$, rather it remains stable for the network types considered here. In the case of the deterministically-constructed outerplanar hierarchical networks, the mean period plateaus at approximately $\langle T \rangle=20$ time steps for successive generations. Overall, the mean periods are highest for scale-free networks. We see that larger Erd\"os-R\'enyi networks (with same edge probability $p=0.07$) tend to have smaller mean periods. We expect the time-scale for higher $N$ to remain similarly bounded.

\subsection{Open-system PageRank via continuous-time quantum walk}

The continuous-time quantum walk was originally proposed by Farhi and Gutmann out of a study of computational problems reformulated in terms of decision trees \cite{farhi1998quantum}. Following the Schr\"odinger equation, such evolution is described by
\begin{equation}
\frac{d\ket{\Psi(t)}}{dt}=-i\hat{H}\ket{\Psi(t)}
\label{eq:OS1},
\end{equation}
where $\hat{H}$ is the transition rate matrix. Requiring unitary evolution operators in quantum mechanics implies that $\hat{H}$ must be Hermitian, which is generally not the case for a directed walk. To introduce directionality into CTQWs, we employ the open system method using the Lindblad-von Neumann equation, which accounts for the non-unitary nature of the directed walk through coupling with an external environment.

To work with open quantum systems, the concept of a density operator is used as a substitution for wave functions in quantum mechanics. The density operator for the system is defined by \cite{nielson2010quantum}
\begin{equation}
\label{eq:OS2}
\rho=\sum\limits_{i=1}^{N}p_i\ket{\Psi_{i}}\bra{\Psi_{i}}
\end{equation}
where $p_i$ are constants that represent how much of state $\ket{\Psi_{i}}$ is in the final mixed state, with $\sum\limits_{i=1}^{N} p_i=1$.

The Lindblad-von Neumann equation describes how a quantum system evolves after tracing out the environment, and can be written in the form \cite{rivas2012open}:
\begin{equation}
\frac{d\rho}{dt}=-i\hbar[\hat{H},\rho]+\sum\limits_k\gamma_k(\hat{L}_k\rho\hat{L}_k^\dagger-\frac{1}{2}\{\hat{L}_k\hat{L}_k^\dagger,\rho\})
\label{eq:OS4},
\end{equation}
where $\hat{L}_k$ are unitary operators on the space that $\rho$ is in. The set of all $\hat{L}_k$ forms a basis for this space. The matrix $\gamma$ describes how non-energy conserving phenomena such as temperature affect the system. 

To parameterise interpolation between classical (undirected) and classical (directed) behaviours, a damping parameter $\beta$ is introduced into the Lindblad-von Neumann equation:
\begin{equation}
\frac{d\rho}{dt}=-i(1-\beta)[\hat{H},\rho]+\beta\sum\limits_{i,j}\gamma_{ij}(\hat{L}_{ij}\rho\hat{L}_{ij}^\dagger-\frac{1}{2}\{\hat{L}_{ij}\hat{L}_{ij}^\dagger,\rho\})
\label{eq:OS5},
\end{equation}
where $\hat{H}$ is just the adjacency matrix $C$ with no allowance for direction or weighting of the underlying graph. $\gamma$ is taken as the patched connectivity matrix $E$ as per Eq. \eqref{eq:matE} -- this a specific case of the Google matrix $G$ in Eq. \eqref{eq:matG} with $\alpha = 1$. Our approach is equivalent to the quantum PageRank algorithm developed by S\'anchez-Burillo \textit{et al.} in \cite{sanchez2012quantum}, where they set $\gamma=G$ with $\alpha = 0.9$ instead.

We solve the master equation via an eigen-operator method used by Saalfrank \cite{Saalfrank1998open}. This is a linearisation method that turns a non-linear equation into a linear one, whereby Eq. \eqref{eq:OS5} becomes
\begin{equation}
\frac{d\rho}{dt}=-i(1-\beta)\mathcal{L}_H\rho+\beta\mathcal{L}_D\rho=\mathcal{L}_{SO}\rho
\label{eq:OS6}.
\end{equation}
This takes the eigen-operators $\mathcal{L}_H$ and $\mathcal{L}_D$ from the original $N$-dimensional space to a space of $N^2$ dimensions, and the density matrix is vectorised in this set-up.

As per Eq. \eqref{eq:OS6}, $\mathcal{L}_H$ and $\mathcal{L}_D$ can be combined into one operator $\mathcal{L}_{SO}$. For time-independent $\mathcal{L}_{SO}$, this form is readily solved for any time $t$ by taking the matrix exponential of $\mathcal{L}_{SO}$, i.e.
\begin{equation}
\rho=\rho_0e^{\mathcal{L}_{SO}t}
\label{eq:OS7}.
\end{equation}
Convergence to a stationary result for large enough $t$ is guaranteed \cite{sanchez2012quantum}, upon which the occupation probabilities of each node indicate the open-system quantum PageRank, namely
\begin{equation}
I_{OS}(P_i) := \langle i|\rho|i\rangle= \rho_{ii}
\label{eq:OS10}.
\end{equation}

\section{Results}
\label{sec:Results}

To recap, the four PageRank measures considered here are:
\begin{itemize}
\item $I_{cl}$ \eqref{eq:CPR} -- classical Google PageRank
\item $I_{TA}$ \eqref{eq:SG8} -- DTQW-based PageRank using the time-average of the instantaneous quantum PageRank $I_q$
\item $I_{P_{max}}$ \eqref{eq:SG9} -- DTQW-based PageRank using the maximum $I_q$ reached
\item $I_{OS}$ \eqref{eq:OS10} -- CTQW-based PageRank using the open system method
\end{itemize}
We set $\alpha = 0.85$ in the Google matrix $G$ \eqref{eq:matG} for $I_{cl}$, $I_{TA}$, and $I_{P_{max}}$. For $I_{OS}$, we use $\alpha = 1$ in $G$ and $\beta = 0.85$ in the master equation \eqref{eq:OS5}.

For all numerical results below, the DTQW-based PageRank measures $I_{TA}$ \eqref{eq:SG8} and $I_{P_{max}}$ \eqref{eq:SG9} are computed using  $t_{max}=2\langle T_{q}^{5} \rangle$ steps of the evolution operator $\hat{U}^{2}$, with values for $\langle T_{q}^{5} \rangle$ in Table \ref{table:T}. For $I_{OS}$, we use sufficiently large $t$ for convergence to the stationary result.

\subsection{Three types of directed networks considered}

To allow meaningful comparison between the four PageRank measures on various networks, we normalise the PageRanks by dividing through by the maximum value obtained so that the most central node has a PageRank value of 1.

\subsubsection{Outerplanar hierarchical networks}

As introduced by Comellas and Miralles~\cite{comellas2009modeling}, outerplanar hierarchical networks are a family of modular, self-similar, small-world graphs with zero clustering. This family mirrors social, technological, and biological systems with a low clustering \cite{comellas2009vertex}. Outerplanarity refers to the network having an embedding where all nodes lie on the boundary of the exterior face; whereas the hierarchical structure is realised by using a recursive method of network construction.
Each generation $n$ has $N=2^{n+1}$ nodes; thus the network doubles in size for successive $n$, with newly-added nodes being those indexed $2^{n} < i \leq 2^{n+1}$. We follow \cite{paparo2013quantum} in giving directions to the edges.

\begin{figure}[tbh]
	\centering
	\vspace{0cm} \hspace{0cm} \scalebox{0.7}{\includegraphics{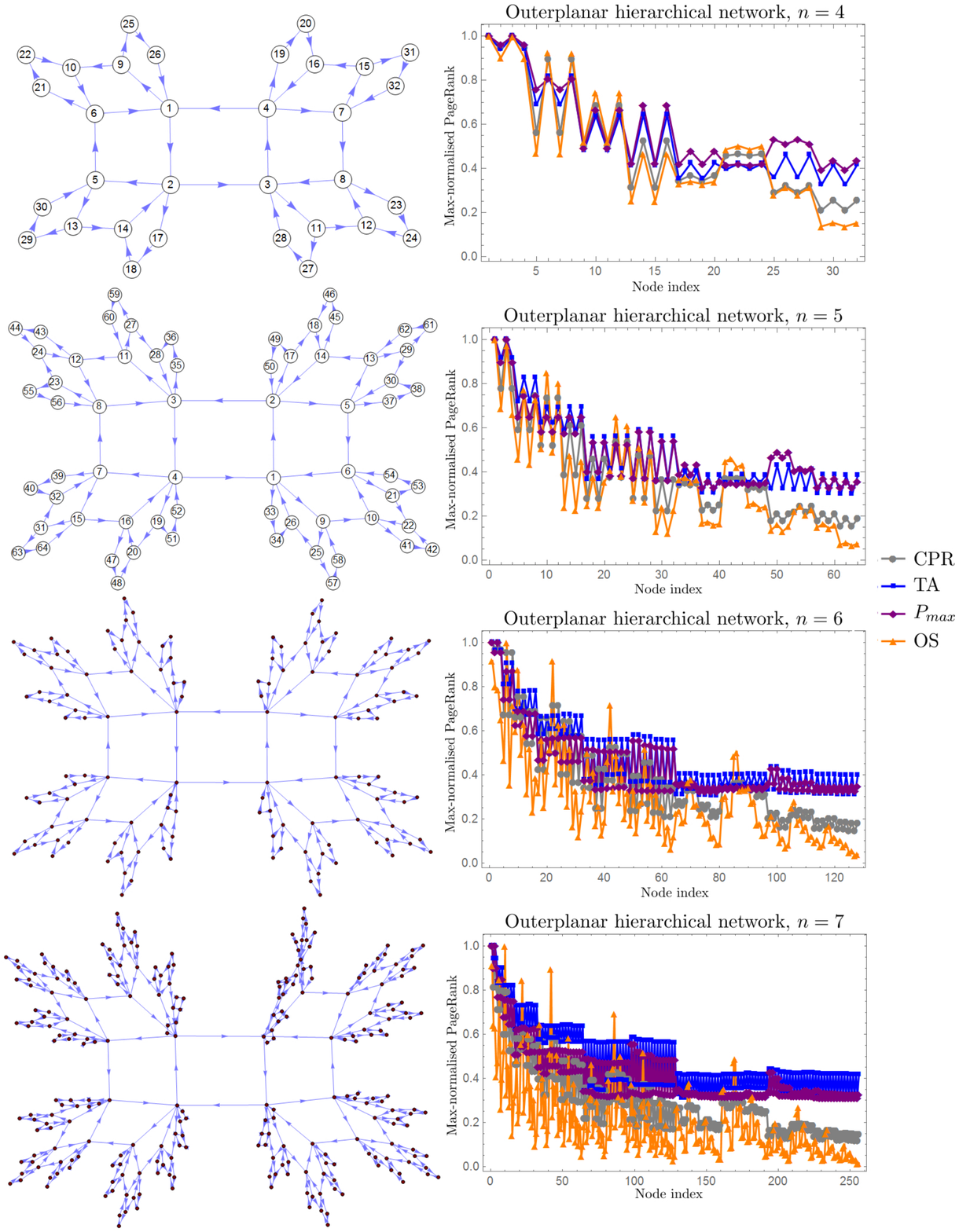}}
\caption{\textbf{PageRanks on outerplanar hierarchical networks for generations \boldmath$n=4, 5, 6, 7$.} $I_{cl}$, $I_{TA}$, $I_{P_{max}}$, and $I_{OS}$ -- normalised by their maximum values -- are plotted in gray, blue, purple, and orange respectively.}
\label{fig:I-OPn}
\end{figure}

As shown in Figure \ref{fig:I-OPn}, all three quantum PageRank measures have a step-like behaviour that reflects the network's hierarchical structure. Within the same hierarchical level, edge directionality gives rise to non-degenerate PageRank values. This intra-level non-degeneracy has a smaller amplitude for the newly-added nodes (evidenced by nodes 17-32 in $n=4$, nodes 33-64 in $n=5$, nodes 65-128 in $n=6$, and nodes 129-256 in $n=7$), thus enabling the PageRank measures to distinguish between pre-existing and newly-added nodes in successive generations.

\begin{figure}[tbh]
	\centering
	\vspace{0cm} \hspace{0cm} \scalebox{0.57}{\includegraphics{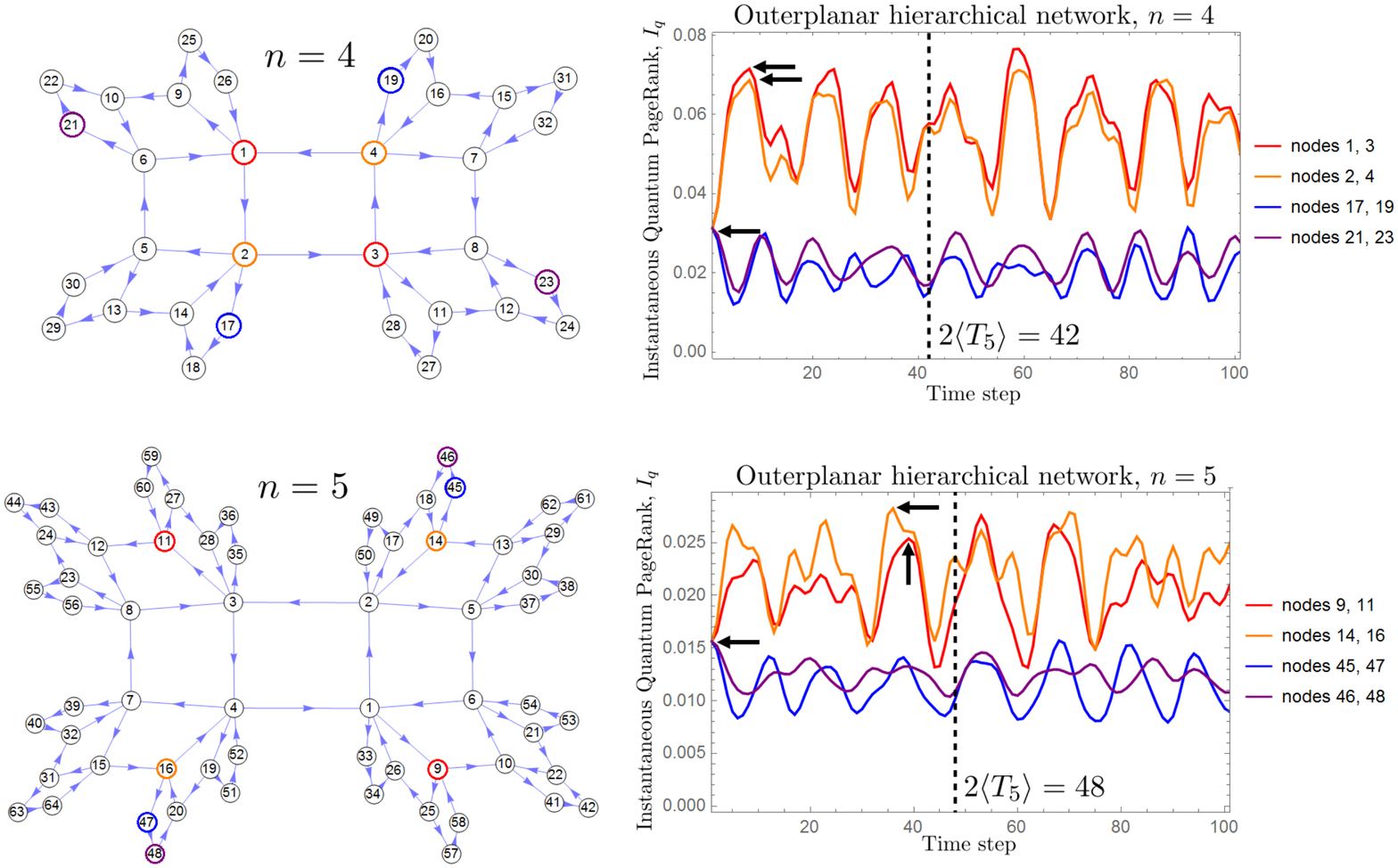}}
\caption{\textbf{Time evolution of $I_q$ on \boldmath$n=4$ and \boldmath$n=5$  outerplanar hierarchical networks.} Left: The selected like-coloured nodes are automorphically equivalent, and have $I_q$ that evolve identically over time. Right: Time evolution of $I_q$. Vertical dashed lines bound $t_{max}=2\langle T_{q}^{5} \rangle$ for the network, within which we determine $I_{P_{max}}$ as indicated by arrows, and $I_{TA}$ by taking the time-averaged probabilities.}
\label{fig:I-OPn45d}
\end{figure}

Comparing $I_{TA}$ and $I_{P_{max}}$, the latter measure tends to give more closely-valued PageRanks within a hierarchical level, particularly for the newly-added nodes. As depicted in Figure \ref{fig:I-OPn45d}, nodes whose oscillatory $I_q$ never peak above the value subsequent to the initial state receive near-similar $I_{P_{max}}=I_q(P_i, t=1)$ after one time step. For example, $I_{P_{max}}$ almost plateaus for nodes 21-24 in $n=4$ and nodes 41-48 in $n=5$. Taking the time average resolves this degeneracy as each node's $I_q$ evolves slightly differently in time.

Furthermore and as would be expected, we see that pairs of automorphically equivalent nodes have identical time evolution of $I_q$. (Two nodes $a$ and $b$ are said to be automorphically equivalent if there exists an isomorphism in which the labels of $a$ and $b$ are interchanged \cite{everett1985role, borgatti1992notions}.) Such nodes are thus ranked identically by $I_{TA}$ and $I_{P_{max}}$, preserving the equivalence also present in $I_{cl}$.
On the other hand, this identical ranking of automorphically equivalent nodes is not preserved by $I_{OS}$. We instead observe a falling pattern of $I_{OS}$ values within the hierarchical levels.

In summary, the four PageRank measures considered here are able to uncover the hierarchical structure present in each network, but disagree on the relative rankings of these levels. Here $I_{OS}$ most closely resembles $I_{cl}$ in ranking each level; but, especially for higher $n$, its intra-level behaviour is in stark contrast to the other measures.

\subsubsection{Scale-free networks}

A power-law degree distribution typifies scale-free networks, that is, the probability $P(k)$ that a node is connected to $k$ other nodes decays as $P(k) \sim k^{-\gamma}$ \cite{barabasi1999emergence}. Such behaviour was first observed by Albert \textit{et al.} in their analysis of the topology of the World-Wide Web, in that the numbers of incoming and outgoing hyperlinks of a webpage both follow a power law over several orders of magnitude \cite{albert1999internet}. Numerous other real-world networks have since been found to be scale-free, from functional networks in the brain \cite{eguiluz2005scale} and protein interactions in cells \cite{albert2005scale}, to social networks and their technical derivatives such as the Internet, e-mail networks, and business collaboration \cite{hein2006scale}.

Inherently linked to the scale-free property of a network is its evolution over time \cite{barabasi2009scale}, which can be modelled by preferential attachment \cite{barabasi2000scale}. Starting with a small number of nodes, new nodes are added with a higher probability of being connected to pre-existing nodes that are already well-connected -- conceptually, ``the rich get richer.'' In this study, we perform PageRank analysis on Bollob\'as \textit{et al.}'s scheme for directed scale-free networks \cite{bollobas2003directed} as implemented in NetworkX \cite{schult2008exploring}. This scheme allows multiple edges and loops.

\begin{figure}[tbh]
	\centering
	\vspace{0cm} \hspace{0cm} \scalebox{0.7}{\includegraphics{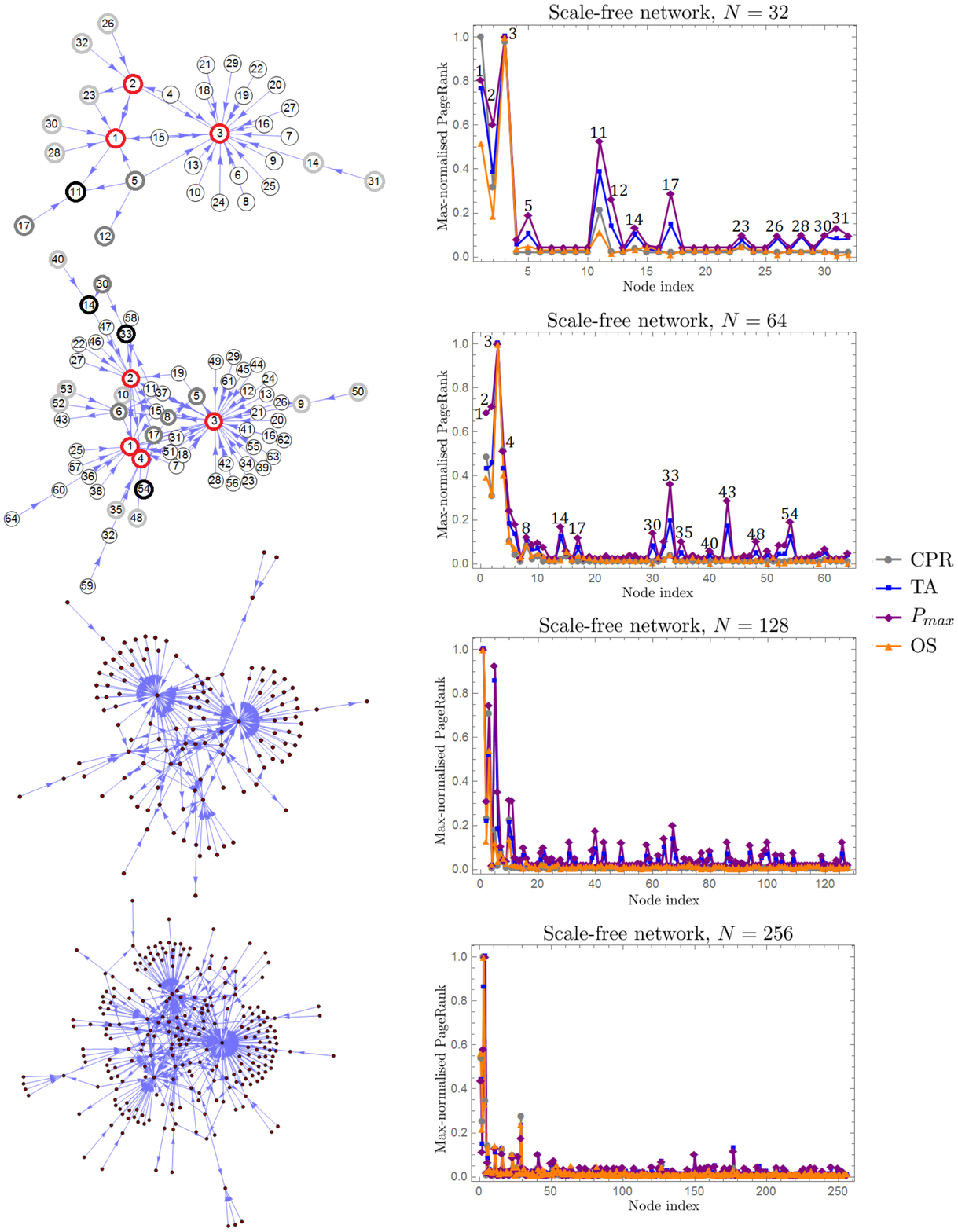}}
\caption{\textbf{PageRanks on directed scale-free networks of sizes \boldmath$N=32, 64, 128, 256$.} $I_{cl}$, $I_{TA}$, $I_{P_{max}}$, and $I_{OS}$ -- normalised by their maximum values -- are plotted in gray, blue, purple, and orange respectively. For the $N=32$ and $N=64$ cases, nodes marked in red are the most central nodes, or hubs; secondary hubs identified by $I_{TA}$ and $I_{P_{max}}$ are marked in grayscale.}
\label{fig:I-SFns}
\end{figure}

\begin{figure}[tbh]
	\centering
	\vspace{0cm} \hspace{0cm} \scalebox{0.6}{\includegraphics{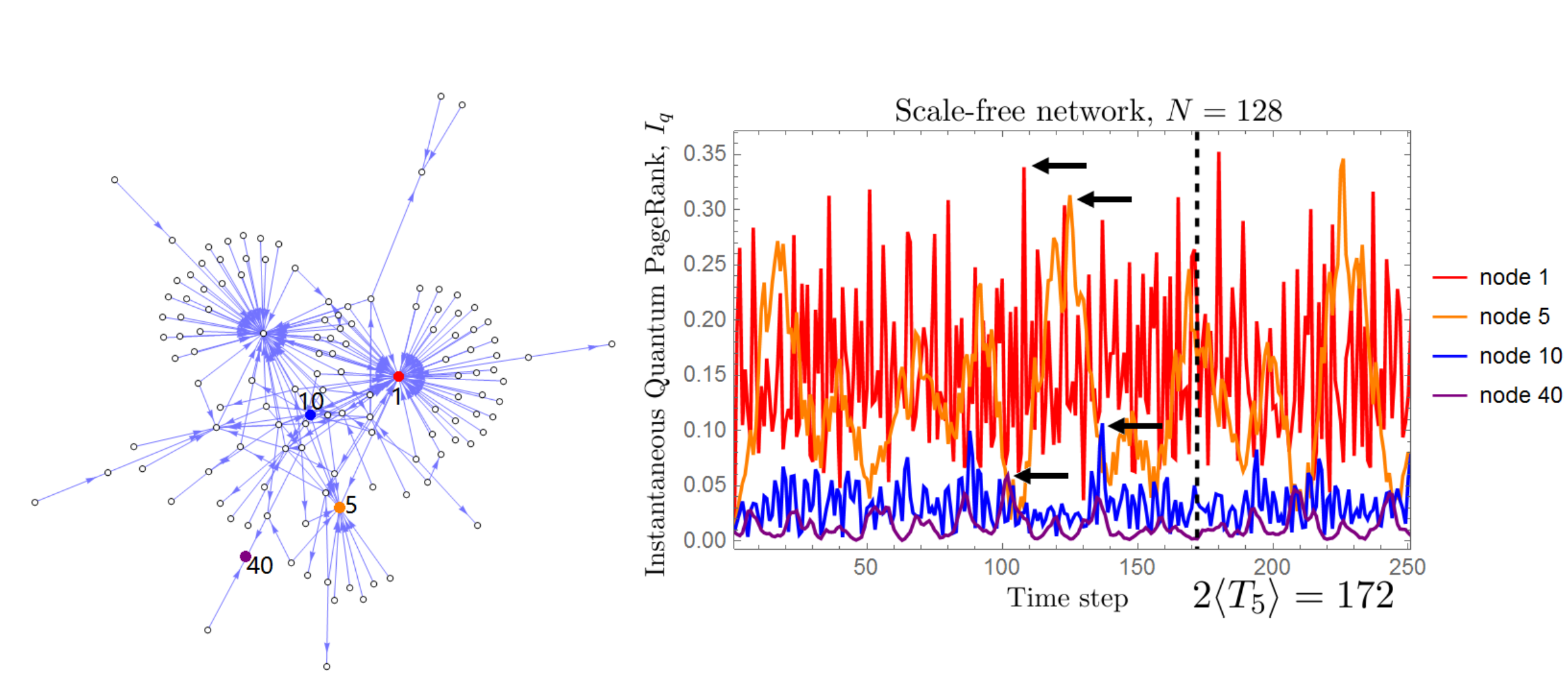}}
\caption{\textbf{Time evolution of $I_q$ on a directed scale-free network of size $N=128$.} The black vertical dashed line bounds $t_{max}=2\langle T_{q}^{5} \rangle$ for the network, within which we determine $I_{P_{max}}$ as indicated by arrows, and $I_{TA}$ by taking the time-averaged probabilities.}
\label{fig:I-SFd}
\end{figure}

Based on Figure \ref{fig:I-SFns}, all PageRank methods largely agree on identifying the most central nodes, or hubs. Figure \ref{fig:I-SFd} provides an example case for obtaining $I_{TA}$ and $I_{P_{max}}$ based on the time evolution of $I_q$.

As previously observed and noted in \cite{paparo2013quantum}, $I_{TA}$ better highlights the secondary hubs compared to $I_{cl}$ rather than concentrating all the importance on the most important hubs. We see that this improved ranking capability of $I_{TA}$ over classical PageRank is even more strongly featured in $I_{P_{max}}$ on both types of scale-free network. We investigate this further in Subsection \ref{subsec:II}.

For the examples considered here, nodes with little or no in-degree receive the lowest importances according to $I_{cl}$ and $I_{OS}$, forming a near-plateau of low-lying nodes.

$I_{TA}$ and $I_{P_{max}}$ are additionally affected by which nodes a given node is pointing to. In the case of the 32-node scale-free network in Figure \ref{fig:I-SFns}, node 17 has unexpectedly high $I_{TA}$ and $I_{P_{max}}$ by virtue of its pointing to secondary hub node 11. Similarly, the nodes pointing to central hubs nodes 1 and 2 have low $I_{cl}$ and $I_{OS}$, but have improved $I_{TA}$ and $I_{P_{max}}$. The more nodes there are that point to a given hub, the less advantage those nodes gain, as is the case for those surrounding node 3. This accords with the observation made by \cite{sanchez2012quantum} for their open-system-quantum-walk-based PageRank, in that nodes connected to hubs distributing their influence amongst a large number of connections receive lower rankings.

Therefore, low-lying nodes with degenerate $I_{cl}$ can be distinguished by $I_{TA}$ and $I_{P_{max}}$ as these measures are more sensitive to the nodes' positions in the network. In particular, a node's ranking is increased by being linked to a network hub, provided it is among a few neighbours of the hub. This ability of quantumness to resolve classical PageRank degeneracies among peripheral nodes is also noted in \cite{paparo2013quantum} and \cite{sanchez2012quantum}.

Finally we observe that $I_{OS}$ closely resembles $I_{cl}$ on these networks. Considering that $I_{OS}$ as implemented here uses damping parameter $\alpha=1 \Rightarrow G=E$ as the matrix coding classical behaviour (directionality) into the walk, this suggests that $\beta$ in the master equation \eqref{eq:OS6} (parameterising the classicality of the walk, here set to $\beta=0.85$) can play a similar role to that of $\alpha$ in classical PageRanking (parameterising random hops to any node, set to $\alpha=0.85$ for $I_{cl}$).

\subsubsection{Erd\"os-R\'enyi random networks}

An Erd\"os-R\'enyi random network of $N$ nodes is constructed by choosing, with common edge probability $p$, whether or not to connect pairs of nodes, with the choices being independent for each node pair \cite{gilbert1959random, erdos1960evolution}. For large $N$, its degree distribution follows a Poisson distribution $P(k)=e^{-\langle k \rangle} \frac{\langle k \rangle^k}{k!}$, where $\langle k \rangle$ is the mean degree. Despite having random edge positions, such a network is rather homogeneous as most nodes have the same degree \cite{albert2002statistical}. In this study, we use directed Erd\"os-R\'enyi random networks generated using NetworkX \cite{schult2008exploring}.

\begin{figure}[tbh]
	\centering
	\vspace{0cm} \hspace{0cm} \scalebox{0.7}{\includegraphics{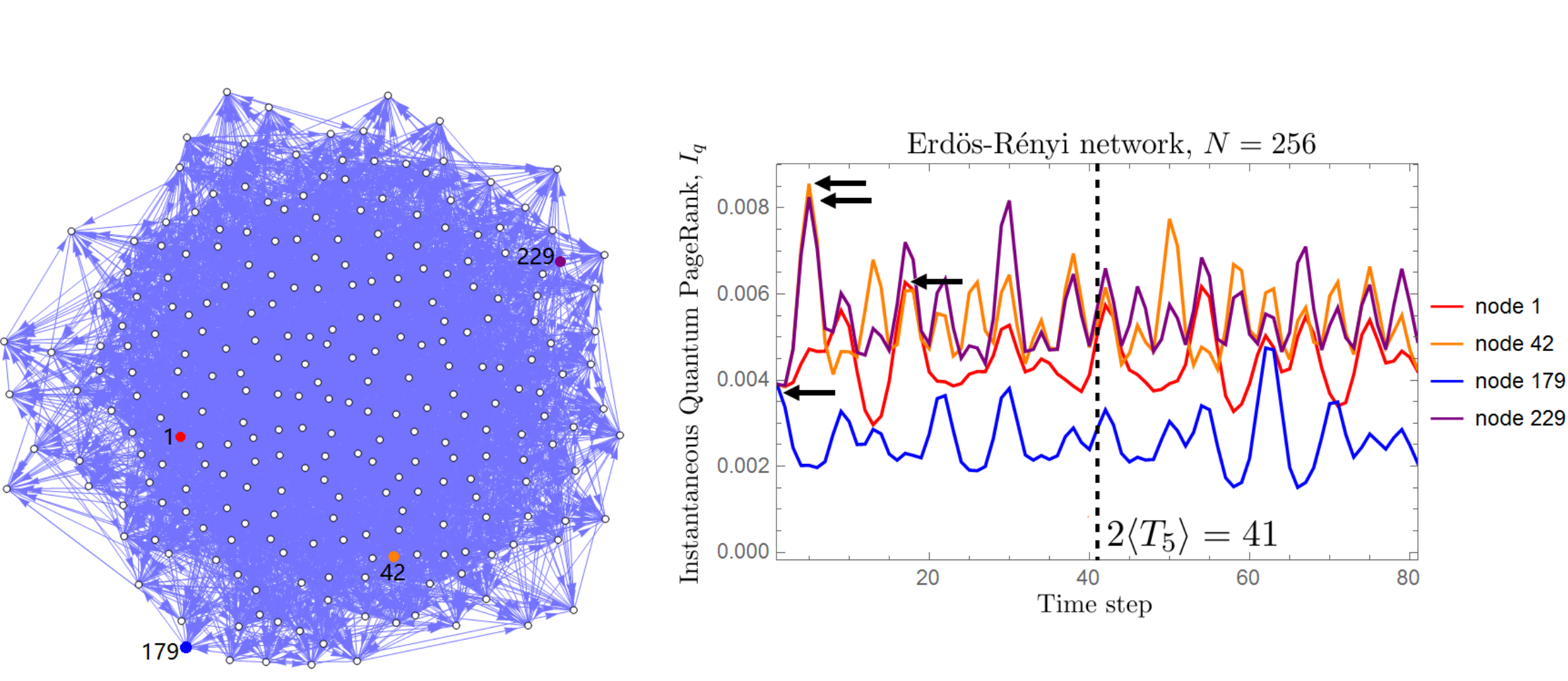}}
\caption{\textbf{Time evolution of $I_q$ on a directed Erd\"os-R\'enyi network of size $N=256$.} The vertical dashed line indicates $t_{max}=2\langle T_{q}^{5} \rangle$ for the network, within which we determine $I_{P_{max}}$ as indicated by arrows, and $I_{TA}$ by taking the time-averaged probabilities.}
\label{fig:I-ERd}
\end{figure}

\begin{figure}[tbh]
	\centering
	\vspace{0cm} \hspace{0cm} \scalebox{0.7}{\includegraphics{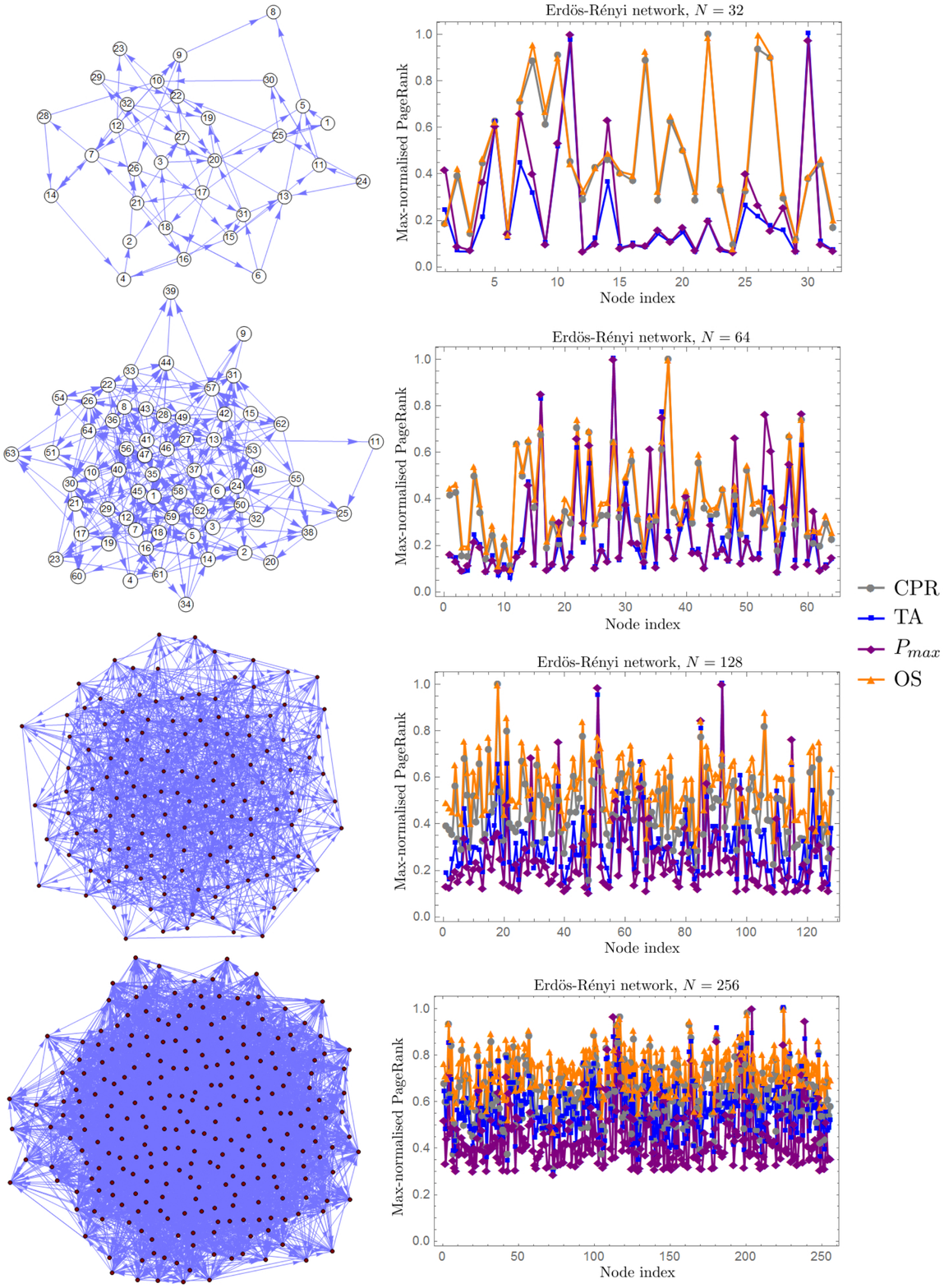}}
\caption{\textbf{PageRanks on directed Erd\"os-R\'enyi random networks of sizes \boldmath$N=32, 64, 128, 256$.} $I_{cl}$, $I_{TA}$, $I_{P_{max}}$, and $I_{OS}$ -- normalised by their maximum values -- are plotted in gray, blue, purple, and orange respectively.}
\label{fig:I-ERn}
\end{figure}

An example case of obtaining the DTQW-based $I_{TA}$ and $I_{P_{max}}$ is presented in Figure \ref{fig:I-ERd}. Overall results in Figure \ref{fig:I-ERn} demonstrate that for each PageRank measure, most nodes receive similar PageRanks with no discernible hubs. This is an ostensibly different distribution of PageRanks in contrast to the directed scale-free networks analysed earlier. Subsection \ref{subsec:III} further studies this localised/delocalised behaviour of the walker on these two network types.

In summary, consistent with our analysis on scale-free networks, we find that $I_{P_{max}}$ provides an alternative measure to $I_{TA}$ as both exhibit similar features; whereas $I_{OS}$ more closely resembles $I_{cl}$.

\subsection{Detection of secondary hubs on scale-free graphs}
\label{subsec:II}

The preliminary PageRank analyses on scale-free graphs suggest that the quantum-walk-based PageRank measures tend to give higher ranks to secondary hubs compared to classical PageRank. To quantify such secondary-hub detection for each PageRank scheme, we analyse an ensemble of 30 directed scale-free networks of size $N=256$ by categorising the nodes according to their PageRanks.

As before, for each network and PageRank method, we normalise the nodes' PageRanks by dividing through by the maximum so that the most central node has a PageRank of 1. Let the mean of these maximum-normalised PageRanks be $a$. Then we classify a node with PageRank $x$ as a
\begin{itemize}
\item main hub if $x \geq c a$,
\item secondary hub if $a \leq x < c a$,
\item low-importance node hub if $x < a$,
\end{itemize}
where $c>1$ is a fixed constant. Here we choose $c=10$.

\begin{figure}[tbh]
	\centering
	\vspace{0cm} \hspace{0cm} \scalebox{0.56}{\includegraphics{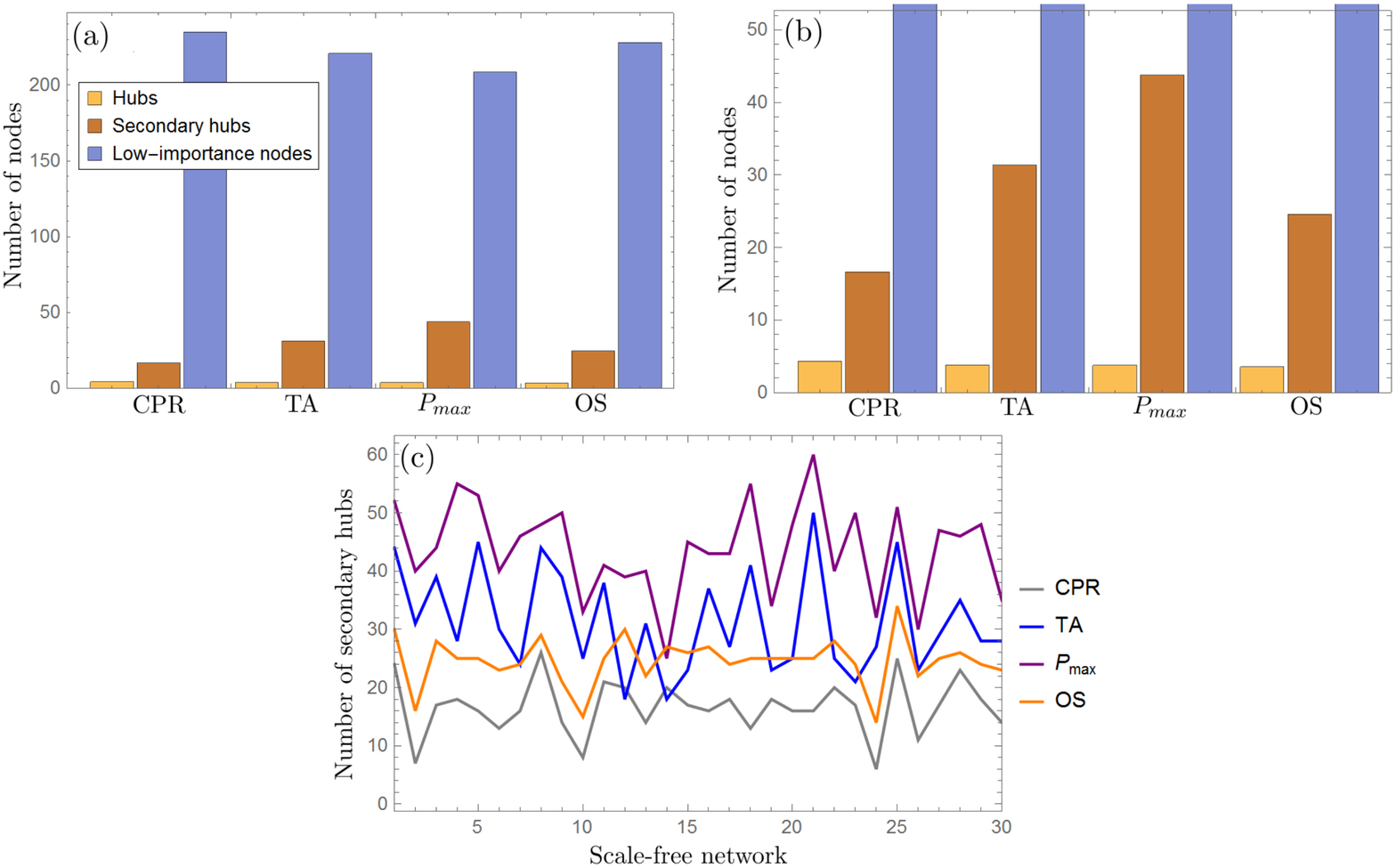}}
\caption{\textbf{Secondary hub resolution by PageRank algorithms on an ensemble of 30 directed scale-free networks of size \boldmath$N=256$.} (a) A histogram of nodes classified as main hubs, secondary hubs, or low-importance nodes based on their (from left to right) $I_{cl}$, $I_{TA}$, $I_{P_{max}}$, and $I_{OS}$ values respectively. (b) Zooming into (a). The quantum PageRanks $I_{TA}$, $I_{P_{max}}$, and $I_{OS}$ respectively identify approximately 1.9, 2.6, 1.5 times more secondary hubs than $I_{cl}$. (c) The number of secondary hubs as measured by $I_{cl}$ (gray), $I_{TA}$ (blue), $I_{P_{max}}$ (purple), and $I_{OS}$ (orange) for each of the scale-free networks in the ensemble.}
\label{fig:II-SF}
\end{figure}

Results plotted in Figure \ref{fig:II-SF} affirm our earlier observations on scale-free networks. Comparing all Pagerank measures considered, $I_{P_{max}}$ identifies the most number of nodes as secondary hubs, followed by $I_{TA}$, $I_{OS}$, and finally $I_{cl}$.

Based on Figure \ref{fig:II-SF}(c), the set of nodes classified as secondary hubs by $I_{TA}$ is most likely a subset of those identified by $I_{P_{max}}$. In particular, $I_{P_{max}}$ never picks out less secondary hubs than $I_{TA}$; moreover both largely follow the same trend in quantifying secondary hubs across the network ensemble. On the other hand, $I_{OS}$ exhibits a similar trend to $I_{cl}$. Although not as stark a difference as the other two quantum measures, $I_{OS}$ also outperforms $I_{cl}$ in secondary hub detection.

\subsection{Localisation-delocalisation transition}
\label{subsec:III}

To further compare the abilities of the various PageRank schemes to distinguish between scale-free and Erd\"os-R\'enyi random networks, we study their localisation behaviour on these network types. Since each PageRank scheme corresponds to a classical or quantum walk along nodes, we can infer the walker's degree of localisation based on the PageRank distribution across the network.

Within scale-free networks, the presence of hubs -- nodes of unusually high degree -- is the fundamental cause of localisation. Such localisation poses a problem for conventional eigenvector centrality because most of the weight of the centrality concentrates on a small number of nodes, thus necessitating the random-hop term in the Google matrix  \cite{martin2014localization}. On the other hand, the relatively homogeneous degree distribution in Erd\"os-R\'enyi random networks is expected to favour a delocalised phase of the walker, as observed in \cite{paparo2013quantum}.

To quantify localisation, we use the Inverse Participation Ratio (IPR) defined as

\begin{equation}
\xi := \sum_{i=1}^{N}\left[ \text{Pr}(X=i) \right]^{2 r}
\label{eq:IPR},
\end{equation}
where $r>0$ is a freely-chosen integer, and is fixed.

As in \cite{paparo2013quantum}, we appropriate $\xi$ to the case of PageRanks by considering $X$ as a random variable whose realisations are the nodes of the network. Then the probability $Pr(X=i)$ corresponds directly to node $i$'s PageRank $I_{cl}$, $I_{TA}$, or $I_{OS}$ -- for each of these measures, the sum of all PageRanks over the network is one. For a consistent probability interpretation of $I_{P_{max}}$, we divide its values by the mean $\langle I_{P_{max}} \rangle$, so that all such normalised values sum to one.

In the case of complete delocalisation, the walker's probability distribution would be uniform across the network, that is, $\text{Pr}(X=i)=\frac{1}{N} \forall i$. If the walker is fully localised on one node $j$, $\text{Pr}(X=i)=\delta_{i j}$, the Kronecker delta. Hence the IPR for these limiting cases is

\begin{equation}
\xi =
\begin{cases} 1 &\mbox{if the walker is localised} \\ 
N^{1-2 r} & \mbox{if the walker is delocalised.}
\end{cases}
\label{eq:IPRlim}
\end{equation}

Rewrite the IPR as $\xi= N^{-\tau_{2 r}}$ with
\begin{equation}
\tau_{2 r} := (2r-1) + \Delta_{2r}
\label{eq:IPRtau},
\end{equation}
where $\Delta_{2r}$ is the normalised anomalous dimension, which interpolates between $\Delta_{2r}=1-2r$ for a localised phase and $\Delta_{2r}=0$ for a delocalised one. Then
\begin{equation}
\log \xi \sim (1-2 r - \Delta_{2 r}) \log N
\label{eq:IPRlog}.
\end{equation}
Choosing $r=1$ and plotting $\log \xi$ against $\log N$, analytical values for the gradient of the plot are $a=0$ and $a=-1$ for complete localisation and delocalisation respectively.

Based on Eq. \eqref{eq:IPRlog}, we analyse the localisation-delocalisation transition of each PageRank measure. We compute the IPR for scale-free and Erd\"os-R\'enyi networks of size $N=32, 64, 128, 256$, and $512$, using an ensemble of ten graphs for each $N$.

\begin{figure}[tbh]
	\centering
	\vspace{0cm} \hspace{0cm} \scalebox{0.65}{\includegraphics{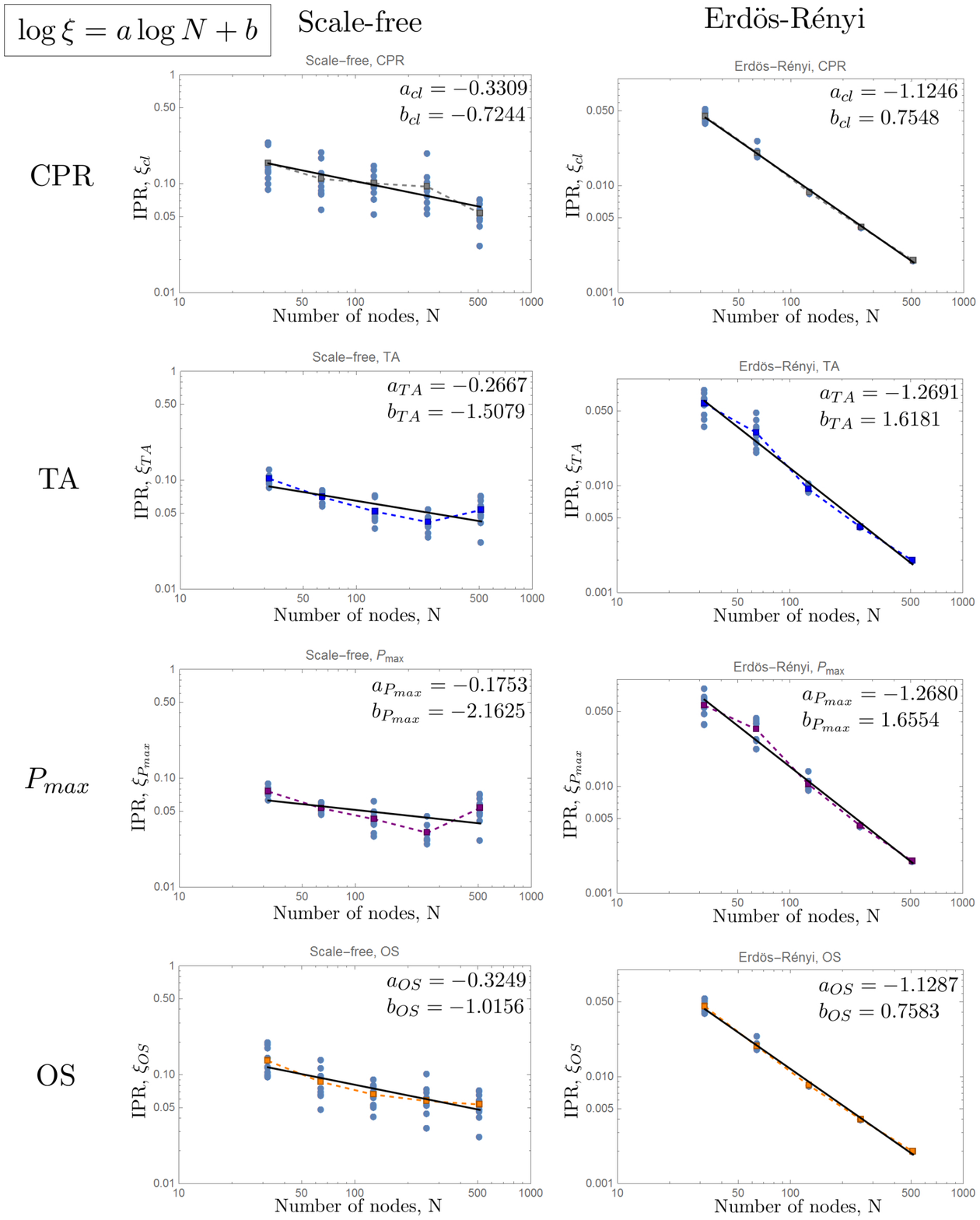}}
\caption{\textbf{Log-log plots of IPR (with \boldmath$r=1$) against the number of nodes in the network for the four PageRank schemes.} Successive rows correspond to $I_{cl}$, $I_{TA}$, $I_{P_{max}}$, and $I_{OS}$; left and right columns correspond to directed scale-free and Erd\"os-R\'enyi networks respectively. An ensemble of ten graphs is used for each network type and size, with a dotted line joining the mean IPR values for each size. Linear model fits are performed according to $\log \xi = a \log N + b$ and plotted in black, with parameter values $a$ and $b$ as indicated.}
\label{fig:III-IPR}
\end{figure}

As shown in Figure \ref{fig:III-IPR}, performing linear fits on the resulting log-log plots yield numerical gradient values $a$ that are near-zero for the scale-free plots, and close to $-1$ for the Erd\"os-R\'enyi plots. These values respectively correspond to localised and delocalised phases of the walker, as was expected based on the presence and absence of hubs in these networks. All four PageRank measures are thus able to distinguish between scale-free and Erd\"os-R\'enyi random networks by virtue of the walker's localisation behaviour.

\subsection{Power law behaviour on scale-free networks}
\label{subsec:IV}

Analyses of classical PageRank on real-world Web graphs have revealed a power law distribution of PageRank values \cite{donato2004large, pandurangan2002using}. This reflects a characteristic property of such scale-free networks, in that only a few main hubs account for much of the PageRank allocation in such scale-free networks. In particular, 
\begin{equation}
I_j \sim j^{-\lambda} 
\label{eq:Plaw},
\end{equation}
where the $I_j$ are the PageRanks of nodes sorted in descending order, and $\lambda$ is the power law scaling coefficient. Such a power law behaviour confirms that the PageRank algorithm is able to reveal a network's scale-free nature, with $\lambda$ measuring the relative importance of hubs with respect to the other less important nodes \cite{paparo2013quantum}.

Here we verify and compare such power law behaviour for the PageRank measures $I_j = I_{cl}, I_{TA}, I_{P_{max}}\text{, and }I_{OS}$. As per Subsection \ref{subsec:III}, the $I_{P_{max}}$ values given by Eq. \eqref{eq:SG9} are normalised to sum to one. For each measure, a plot of the logarithm of the sorted PageRank values $I_j$ against the logarithm of the node index $j$ has slope $\lambda$. We analyse an ensemble of 30 scale-free networks with $N=256$ generated using NetworkX.

\begin{figure}[tbh]
	\centering
	\vspace{0cm} \hspace{0cm} \scalebox{0.56}{\includegraphics{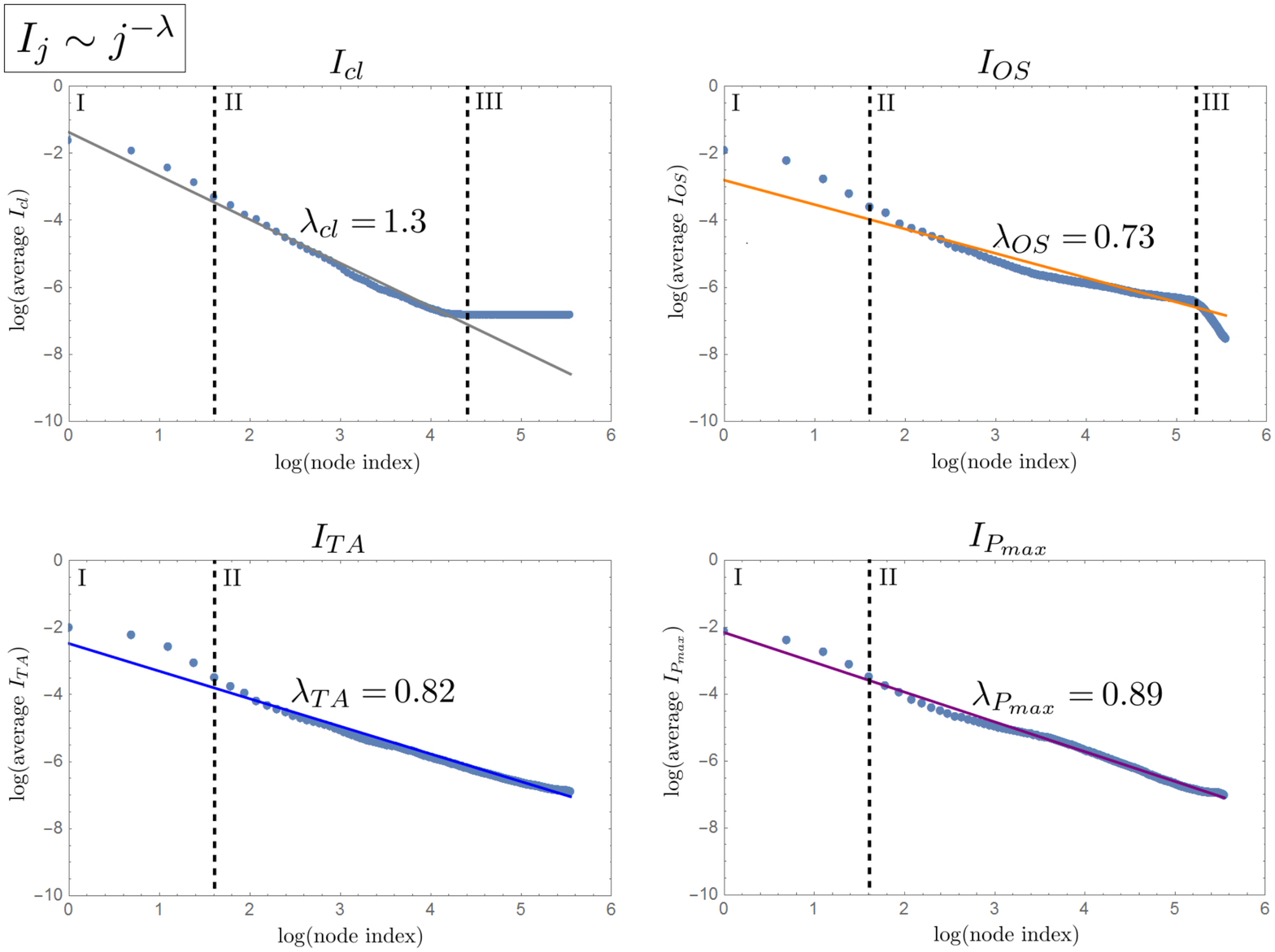}}
\caption{\textbf{Plots of the logarithms of the mean PageRanks for each measure over an ensemble of 30 scale-free networks against the logarithm of the node index (nodes sorted in descending PageRank order).} Particularly over region II, the PageRanks follow a power law distribution $I_j \sim j^{-\lambda} $ across the nodes with fitting parameter $\lambda$. For $I_{cl}$ and $I_{OS}$, nodes in the tail region III are ignored in performing the linear fit.}
\label{fig:IV-all}
\end{figure}

Figure \ref{fig:IV-all} contains plots of the logarithm of the ensemble mean PageRanks against $j$, with linear fits yielding $\beta$ for each PageRank measure. All PageRank measures display a power law behaviour with scaling coefficients $\lambda_{OS} < \lambda_{TA} < \lambda_{P_{max}} < \lambda_{cl}$. The quantum-walk-based measures thus tend to concentrate less importance on the hubs compared to classical PageRank.

As noted in \cite{paparo2013quantum}, the power law behaviour of $I_{TA}$ interpolates over a larger portion of the data compared to $I_{cl}$, marked as region II in Figure \ref{fig:IV-all}. We observe that such a smoother power-law behaviour is preserved by $I_{P_{max}}$. Therefore, $I_{TA}$ and $I_{P_{max}}$ can both better distinguish the low-lying nodes on scale-free networks.

In contrast for $I_{cl}$, and as observed in \cite{pandurangan2002using}, the plot flattens out for nodes in region III with very low PageRank. Such a tail region without the general power-law behaviour is also present for $I_{OS}$, but is characterised instead by a sharp decrease in PageRank. Nodes of lowest importance are thus penalised by $I_{OS}$.

\begin{figure}[tbh]
	\centering
	\vspace{0cm} \hspace{0cm} \scalebox{0.46}{\includegraphics{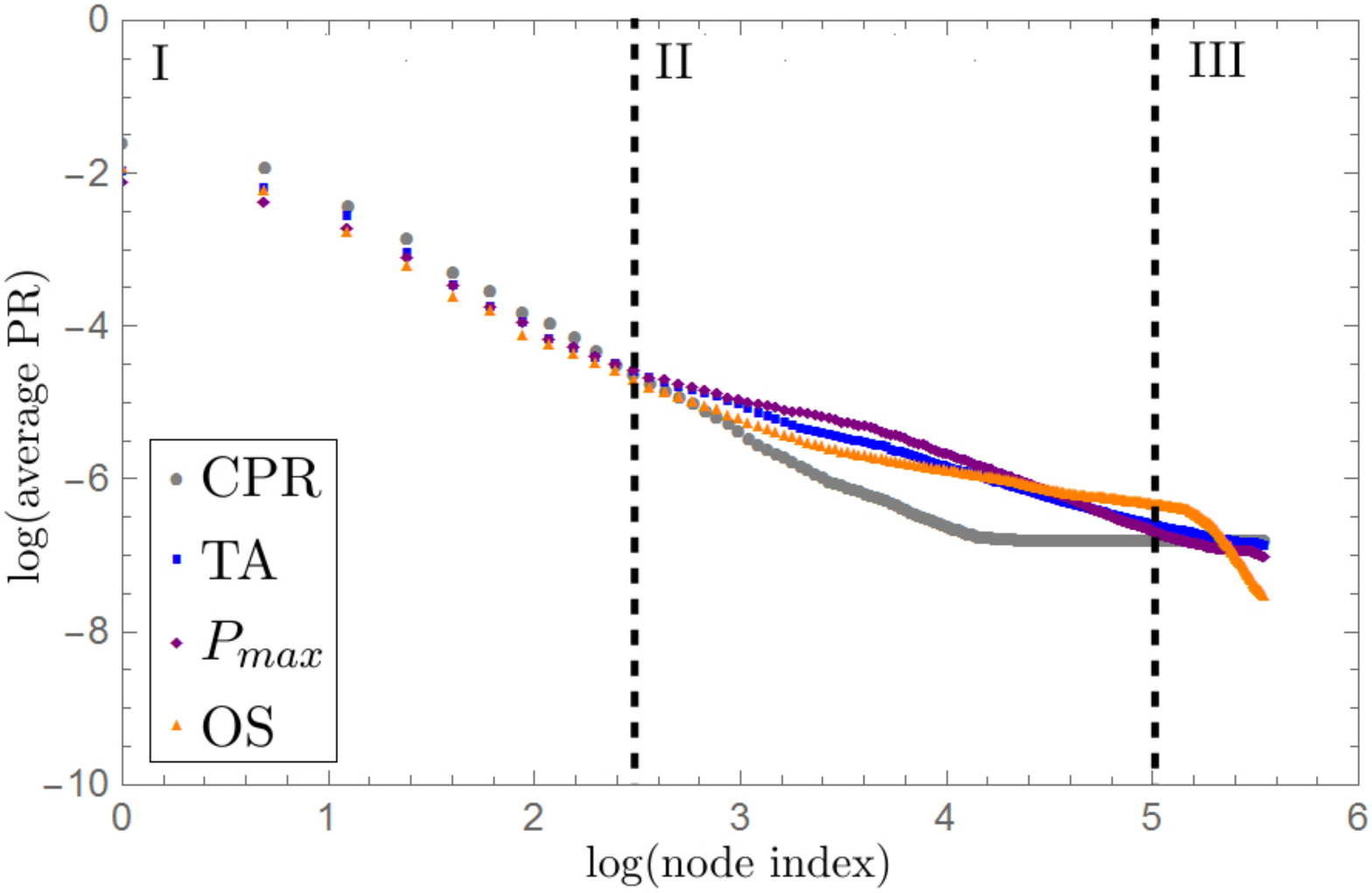}}
\caption{Combined plot of the logarithms of the mean PageRanks from each measure over an ensemble of 30 scale-free networks versus the logarithm of the node index (nodes sorted in descending PageRank order).}
\label{fig:IV-plot}
\end{figure}

Finally, Figure \ref{fig:IV-plot} presents an overall plot of each PageRank measure's distribution on scale-free networks. As shown in region II of the plot, intermediate nodes (including the secondary hubs discussed in Subsection \ref{subsec:II}) are given higher ranks according to (in descending order) $I_{P_{max}}$, $I_{TA}$, and $I_{OS}$ compared to $I_{cl}$. To compensate for this improved ranking, main hubs in region I receive lower quantum than classical PageRanks.

\section{Discussion and conclusions}
\label{sec:Discussion and conclusions}

In this article, we have presented a comparative investigation of three quantum PageRank measures, two of which are based on a DTQW, while the third uses an open-system CTQW. Extending the work in \cite{paparo2013quantum}, we further studied the periodic nature of the instantaneous quantum PageRank $I_q$. In particular, we utilised it to propose a suitable time-scale $t_{max}=2\langle T_{q}^{5} \rangle$ for the corresponding DTQW based on the periods of the hub nodes. Such a time-scale was observed to not scale with network size, making it feasible even use on for larger networks. In addition to taking $I_{TA}$, we proposed a new measure $I_{P_{max}}$ to extract PageRanks from $I_q$. For the open-system PageRank proposed in \cite{sanchez2012quantum}, we investigate the specific case of $\alpha=1$ and $\beta=0.85$ as defined here.

We have demonstrated that all three quantum PageRank measures are able to distinguish between outerplanar hierarchical, scale-free, and Erd\"os-R\'enyi directed networks. Through a comparative view, we observed similarities in rankings given by the DTQW-based PageRanks $I_{TA}$ and $I_{P_{max}}$, and between the open-system-based PageRank $I_{OS}$ and the classical PageRank $I_{cl}$.

When applied to scale-free networks, the quantum PageRank schemes were better able to highlight secondary hubs in scale-free networks than the classical scheme, which tends to concentrate PageRanks on a few main hubs. This affirmed the results in \cite{paparo2013quantum} and \cite{sanchez2012quantum}; moreover we found that this quantum advantage is most apparent in $I_{P_{max}}$, followed by $I_{TA}$ and $I_{OS}$. 

We used the Inverse Participation Ratio (IPR) to characterise the walker's localisation on scale-free and Erd\"os-R'enyi networks. We showed that for all four PageRank methods, the walker is in a localised phase on the hub-containing scale-free networks, and is delocalised on the hubless Erd\"os-R'enyi random networks. Therefore, each PageRank scheme is shown to clearly distinguish between these two types of networks.

Lastly, the distribution of quantum PageRanks was observed to follow a power-law behaviour on scale-free networks -- a property present in classical PageRank. In particular, we showed that $I_{P_{max}}$ preserves the smoother power-law distribution over a larger portion of nodes as found in $I_{TA}$ \cite{paparo2013quantum} compared to the classical case. We observed that $I_{OS}$ scales according to a power law for most nodes, but displays a sharp drop in average PageRank for the nodes of lowest importance, thus indicating that these nodes are penalised by $I_{OS}$ compared to $I_{cl}$.

In summary, we have shown that over the proposed time scale, $I_{P_{max}}$ does provide a feasible alternative measure to $I_{TA}$. It is interesting to note that $I_{OS}$, constructed here without the random hopping term present in $I_{cl}$ (as $\alpha_{OS}=1$ while $\alpha_{cl}=0.85$ in the Google matrix) but with interpolation between the quantum/undirected and classical/directed behaviours (as $\beta=0.85$), resembles the classical PageRank measure that itself interpolates between random hops and the underlying graph.

In future, it will be instructive to further analyse the oscillation periods of $I_q$ on different network types, particularly for deterministically-constructed graphs. From the cases considered here, we posit that the oscillation periods of $I_q$ depend not on the number of nodes in the network, but on the distribution and density of edges between nodes.

It will be useful to study the effect of the damping parameter $\alpha$ on $I_{P_{max}}$ and $I_{OS}$. In \cite{paparo2013quantum}, $I_{TA}$ was found to be stable over a larger range of $\alpha$ values compared to $I_{cl}$. We expect this higher robustness to be preserved by $I_{P_{max}}$ as it is also derived from $I_q$. Furthermore for $I_{OS}$, it will be useful to study the effects of both parameters $\alpha$ and $\beta$ in the Google matrix and master equation respectively, as we have noted here that $\beta$ seems to replicate the effects of $\alpha$ in $I_{cl}$.

As in \cite{paparo2013quantum}, the sensitivity of the PageRank measures to coordinated attacks on hubs in scale-free networks is worth investigating. We expect $I_{P_{max}}$ to be most affected by the selective removal of hubs, as this scheme gives higher rankings to intermediate nodes compared to the others.

Finally, the efficient implementation of such quantum PageRank schemes remains an open problem. While an efficient quantum-circuit-based implementation is proven to be possible for sparse unitary operators \cite{harrow2009quantum}, the use of the Google matrix in $I_{TA}$ and $I_{OS}$ causes the associated unitary evolution operator $\hat{U}$ to be dense. For the open system case, the matrix exponential $e^{\mathcal{L}_{SO}}$ also becomes dense even for sparse networks.

\section{Acknowledgements}
This work was supported by resources provided by the Pawsey Supercomputing Centre with funding from the Australian Government and the Government of Western Australia. TL thanks the Okinawa Institute of Science and Technology as much background learning was done there under the research internship programme, supervised by Thomas Busch and Chandrashekar Madaiah.

\section{Author Contributions}

TL made the most significant contribution to this work, including the development of methodology, detailed simulation and analysis especially for DTQW, as well as preparation of the manuscript. JT and JR worked on the open-system CTQW formulation, simulation and analysis, MS provided very helpful feedback on network analysis in general, and JW provided direction and guidance.

\newpage

\begin{table}[ht]
\begin{center}
\begin{tabular}{| c | c | c | c |}
  \hline
  Network type & $N$ & $\langle T_{q}^{5} \rangle$ & $\langle T_{q}^{all} \rangle$ \\
  \hline
  & $32$ & $21.0$ & $19.7$ \\ \cline{2-4}
  & $64$ & $24.0$ & $19.7$ \\ \cline{2-4}
  Outerplanar hierarchical & $128$ & $21.4$ & $19.8$ \\ \cline{2-4}
  & $256$ & $20.2$ & $20.5$  \\ \cline{2-4}
  & $512$ & $21.0$ & $20.2$  \\ \hline
  & $32$ & $95.3$ & $109.0$ \\ \cline{2-4}
  & $64$ & $100.2$ & $120.7$ \\ \cline{2-4}
  Scale-free & $128$ & $85.8$ & $118.9$  \\ \cline{2-4}
  & $256$ & $91.0$ & $111.0$ \\ \cline{2-4}
  & $512$ & $99.1$ & $113.3$  \\ \hline
  & $32$ & $37.7$ & $68.3$ \\ \cline{2-4}
  & $64$ & $56.6$ & $64.4$  \\ \cline{2-4}
  Erd\"os-R\'enyi & $128$ & $54.5$ & $72.5$ \\ \cline{2-4}
  & $256$ & $20.4$ & $33.8$  \\ \cline{2-4}
  & $512$ & $11.0$ & $17.5$ \\ \hline
\end{tabular}
\caption{Mean periods $\langle T_{q}^{5} \rangle$ and $\langle T_{q}^{all} \rangle$ of the three network types considered, with number of nodes $N$. We use $t_{max}=2\langle T_{q}^{5} \rangle$ as the number of time steps to obtain $I_{TA}$ and $I_{P_{max}}$ for a given network type and size.}
\label{table:T}
\end{center}
\end{table}

\end{document}